\documentclass[prl,twocolumn,showpacs]{revtex4-1}
\usepackage{amsmath,amssymb,bm,graphicx,hyperref}

\newcommand\+{\dagger}
\newcommand\<{\langle}
\renewcommand\>{\rangle}
\newcommand\0{{\bm{0}}}
\renewcommand\k{{\bm{k}}}
\newcommand\p{{\bm{p}}}
\renewcommand\P{{\bm{P}}}
\newcommand\q{{\bm{q}}}
\newcommand\eps{\varepsilon}
\newcommand\ek{\eps_\k}
\newcommand\ep{\eps_\p}
\newcommand\eP{\eps_\P}
\newcommand\Ep{E_\p}
\newcommand\Eq{E_\q}
\newcommand\eF{\eps_\mathrm{F}}
\newcommand\kF{k_\mathrm{F}}
\newcommand\FS{\mathrm{FS}}
\newcommand\SF{\mathrm{SF}}
\newcommand\A{\mathrm{A}}
\newcommand\T{\mathrm{T}}
\newcommand\AT{\mathrm{AT}}
\newcommand\D{\mathrm{D}}

\begin{document}

\title{Polaronic atom-trimer continuity in three-component Fermi gases}

\author{Yusuke Nishida}
\affiliation{Department of Physics, Tokyo Institute of Technology,
Ookayama, Meguro, Tokyo 152-8551, Japan}

\date{December 2014}

\begin{abstract}
 Recently it has been proposed that three-component Fermi gases may
 exhibit a new type of crossover physics in which an unpaired Fermi sea
 of atoms smoothly evolves into that of trimers in addition to the
 ordinary BCS-BEC crossover of condensed pairs.  Here we study its
 corresponding polaron problem in which a single impurity atom of one
 component interacts with condensed pairs of the other two components
 with equal populations.  By developing a variational approach in the
 vicinity of a narrow Feshbach resonance, we show that the impurity atom
 smoothly changes its character from atom to trimer with increasing the
 attraction and eventually there is a sharp transition to dimer.  The
 emergent polaronic atom-trimer continuity can be probed in ultracold
 atoms experiments by measuring the impurity spectral function.  Our
 novel crossover wave function properly incorporating the polaronic
 atom-trimer continuity will provide a useful basis to further
 investigate the phase diagram of three-component Fermi gases in more
 general situations.
\end{abstract}

\pacs{67.85.Lm, 03.75.Ss, 11.10.St, 74.20.Fg}

\maketitle

\section{Introduction}
Interacting Fermi systems appear across a broad range of physics with
various interaction strengths and understanding of their properties is
of fundamental importance.  When there is a weak attraction between two
components of fermions, the system is unstable toward the formation of
Cooper pairs and becomes a Bardeen-Cooper-Schrieffer (BCS) superfluid.
On the other hand, when the attraction is sufficiently strong, the two
components of fermions form a diatomic molecule which undergoes the
Bose-Einstein condensation (BEC) and the system becomes the superfluid
again.  Because there is no sharp distinction between the condensation
of loosely bound Cooper pairs and tightly bound molecules, the above two
types of superfluids can be smoothly connected with increasing the
attraction.  Indeed, it was shown that the ordinary mean-field wave
function smoothly interpolates the BCS and BEC superfluids which
constitutes the celebrated BCS-BEC crossover theory providing unified
understanding of Fermi superfluids~\cite{Eagles:1969,Leggett:1980}.
Since then the BCS-BEC crossover in two-component Fermi gases has been
the subject of extensive studies~\cite{Wilhelm:2012} and now recognized
as a well-established phenomenon, in particular, because of its
experimental realization with ultracold atoms utilizing Feshbach
resonances~\cite{Regal:2004,Zwierlein:2004,Inguscio:2008}.

\begin{figure}[b]
 \includegraphics[width=0.95\columnwidth]{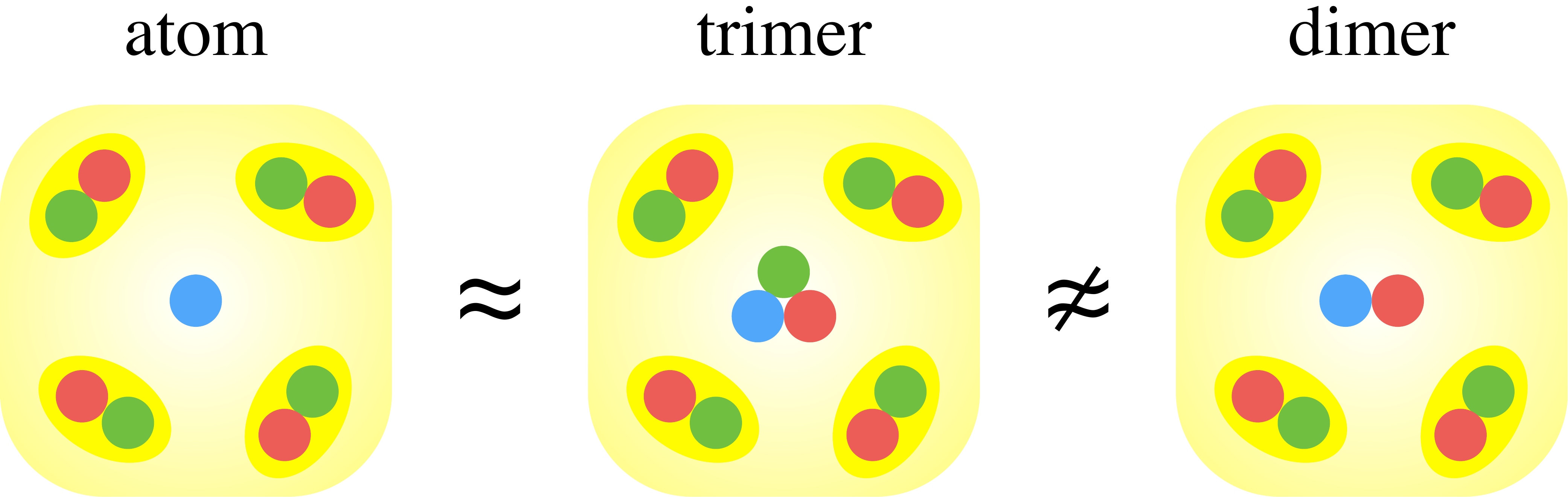}
 \caption{Because of the presence of condensed pairs, atom (left) and
 trimer (middle) are indistinguishable in a superfluid state while they
 are still distinct from dimer (right).  \label{fig:continuity}}
\end{figure}

Yet richer crossover physics may be found in three-component Fermi
gases~\cite{Ottenstein:2008,Huckans:2009,Nakajima:2010}.  When an
attraction between three components of fermions is weak, two of them
form Cooper pairs and condense while there is always one component that
remains unpaired and forms a Fermi
sea~\cite{Honerkamp:2004,Paananen:2006,He:2006,Cherng:2007,footnote}.
Then, with increasing the attraction, loosely bound Cooper pairs will
smoothly evolve into tightly bound molecules according to the BCS-BEC
crossover.  But, what will happen to unpaired fermions?  A new
possibility recently proposed is that unpaired fermions forming a Fermi
sea smoothly change their character from atoms to triatomic molecules
(trimers) with increasing the attraction~\cite{Nishida:2012}.  At first
glance, it may seem surprising and even impossible because an atom and
trimer have different quantum numbers (particle numbers) and thus cannot
be smoothly connected.  However, because the condensation of Cooper
pairs or molecules violates the particle number conservation in units of
two, the atom and trimer are actually indistinguishable in a superfluid
state and thus can be smoothly connected (see
Fig.~\ref{fig:continuity}).  This new type of crossover physics
potentially emerging in three-component Fermi gases is termed an
``atom-trimer continuity'' in analogy with the ``quark-hadron
continuity'' in a superfluid nuclear matter where deconfined quarks with
three colors are considered to smoothly evolve into confined baryons
with decreasing the nuclear density~\cite{Schaefer:1999,Wilczek:2007}.

The atom-trimer continuity was originally inspired by exploring the
universal phase diagram of a three-component Fermi gas in the vicinity
of a narrow Feshbach resonance~\cite{Nishida:2012}.  It was found there
by controlled analyses that the unpaired Fermi sea coexisting with
condensed pairs is composed of atoms in the weak coupling dense limit
while composed of trimers in the strong coupling dilute limit.  Whether
they are actually smoothly connected or not, however, still remains
unestablished because of the lack of a unified theoretical framework
that smoothly interpolates the above two limits.  This difficulty
largely originates from the fact that, unlike the BCS-BEC crossover for
two-component Fermi gases, the ordinary mean-field approximation does
not work for three-component Fermi gases because it completely misses
three-body correlations playing an essential role
here~\cite{Bedaque:2009,Chang:2006,Rapp:2007,Floerchinger:2009,Niemann:2012,Nishida:2012,Nygaard:2014}.
Therefore, it is challenging but highly desired to develop a new
crossover theory suitable for three-component Fermi gases.  Important
progress toward such a goal is to be made in this Letter by studying the
corresponding polaron problem in which a single impurity atom of one
component interacts with condensed pairs of the other two components of
fermions with equal populations.

\section{Crossover wave function}
As emphasized in Ref.~\cite{Nishida:2012}, the BCS-BEC crossover in
two-component Fermi gases is often studied across a broad Feshbach
resonance, where a three-component Fermi gas, however, does not have a
universal many-body ground state because of the Thomas
collapse~\cite{Thomas:1935}.  A minimal extension to cure this problem
is to consider a narrow Feshbach resonance which is characterized by the
nonzero resonance range as well as the $s$-wave scattering
length~\cite{Petrov:2004,Gogolin:2008,Pricoupenko:2013,Hazlett:2012}.
Assuming the same inter-component interaction for all possible three
pairs~\cite{Zhang:2009,O'Hara:2011}, the polaron problem of a
three-component Fermi gas in the vicinity of a narrow Feshbach resonance
is described by the two-channel Hamiltonian,
$H=H_0-\mu\left(N_1+N_2\right)$, which consists of the canonical part
\begin{align}
 H_0 &= \sum_{i=1,2,3}\sum_\p\left[\ep\,\psi_i^\+(\p)\psi_i(\p)
 + \left(\frac\ep2+\nu\right)\phi_i^\+(\p)\phi_i(\p)\right] \notag\\
 & + \frac{g}{2\sqrt{V}}\sum_{i,j,k}\sum_{\P,\p}
 \left[\epsilon_{ijk}\,\phi_i^\+(\P)\psi_j(\p)\psi_k(\P-\p)
 + \mathrm{H.c.}\right]
\end{align}
with $\epsilon_{ijk}$ being the antisymmetric tensor and the particle
number operator for the $i$th component of fermions:
\begin{align}
 N_i = \sum_\p\bigg[\psi_i^\+(\p)\psi_i(\p)
 + \sum_{j\neq i}\phi_j^\+(\p)\phi_j(\p)\bigg].
\end{align}
Here $\psi_i^\+(\p)$ and $\phi_i^\+(\p)$ with $i=1,\,2,\,3$ create
fermionic atoms and bosonic molecules with momentum $\p$ and obey the
usual (anti)commutation relations, while $\ep\equiv\p^2/(2m)$ is the
single-particle kinetic energy with the same mass $m$ assumed for all
three components of fermions.  The chemical potential $\mu$ is
introduced to only the first two components of fermions and chosen so
that each component has the fixed particle number density of
$n_1=n_2\equiv\kF^3/(6\pi^2)$ in the thermodynamic limit $V\to\infty$.
The bare detuning $\nu$ of a molecule $\phi_i$ and its coupling $g$ to
two atoms $\epsilon_{ijk}\psi_j\psi_k$ are related to the $s$-wave
scattering length $a$ and the resonance range $R_*$ by
\begin{align}
 \frac{\nu}{g^2} = -\frac{m}{4\pi a}
 + \int_{|\p|<\Lambda}\!\frac{d\p}{(2\pi)^3}\frac1{2\ep}
\end{align}
and $g^2=4\pi/(m^2R_*)$, where $\hbar=1$ and the momentum cutoff
$\Lambda$ is sent to infinity at the end of calculations.

We now develop a variational approach aimed at a unified theoretical
framework that smoothly interpolates the polaronic atom and trimer in a
superfluid state.  Because the first two components of fermions have
equal populations and attract each other, they form Cooper pairs and
condense.  In order to describe them, we adopt the ordinary mean-field
wave function
\begin{align}\label{eq:superfluid}
 |\SF\> = \exp\!\left[-\frac{|\lambda|^2}2+\lambda\phi_3^\+(\0)\right]
 \prod_\p\left[u_\p+v_\p\psi_1^\+(\p)\psi_2^\+(-\p)\right]|0\>.
\end{align}
Here the variational parameters $\lambda$, $u_\p$, $v_\p$ are determined
so as to minimize the energy expectation value
$E_\SF\equiv\<\SF|H|\SF\>$ under constraints $|u_\p|^2+|v_\p|^2=1$ for
all $\p$ to satisfy the normalization condition $\<\SF|\SF\>=1$.  With
the method of Lagrange multipliers, we find
\begin{align}
 |u_\p|^2 = \frac{\Ep+\ep-\mu}{2\Ep}
\end{align}
and $v_\p=\Delta/(2\Ep u_\p^*)$, where
$\Ep\equiv\sqrt{(\ep-\mu)^2+|\Delta|^2}$ is the quasiparticle energy and
the gap parameter $\Delta\equiv g\lambda/\sqrt{V}$ as well as the
chemical potential $\mu$ simultaneously solves the gap equation
\begin{subequations}\label{eq:mean-field}
\begin{align}\label{eq:gap}
 -\frac{m}{4\pi a} - \frac{m^2R_*\mu}{2\pi}
 = \int\!\frac{d\p}{(2\pi)^3}\left(\frac1{2\Ep}-\frac1{2\ep}\right)
\end{align}
and the particle number density equation
\begin{align}
 \frac{\kF^3}{6\pi^2} = \frac{m^2R_*|\Delta|^2}{4\pi}
 + \int\!\frac{d\p}{(2\pi)^3}\frac{\Ep-\ep+\mu}{2\Ep}.
\end{align}
\end{subequations}
These two coupled mean-field equations qualitatively describe the
BCS-BEC crossover of the condensed pairs with increasing the attraction
$1/a\kF$~\cite{Marini:1998} and, furthermore, become exact in the narrow
resonance limit $R_*\kF\to\infty$~\cite{Gurarie:2007}.

When a single impurity atom of the third component with momentum $\k$ is
added to the above superfluid state (\ref{eq:superfluid}), the simplest
trial wave function to start with is
\begin{align}\label{eq:atom}
 |\A(\k)\> = z_\k\psi_3^\+(\k)|\SF\>.
\end{align}
While this trial wave function seems to describe an atom-like impurity
on top of the superfluid state, it can also be viewed as a trimer-like
impurity at the same time because the particle number in the superfluid
state fluctuates in units of two.  This peculiarity becomes evident by
decomposing the superfluid wave function $|\SF\>$ in Eq.~(\ref{eq:atom})
into a superposition of fixed particle number wave functions:
\begin{align}\label{eq:decomposition}
 |\A(\k)\> \propto \sum_{N=0}^{\infty}\frac{\psi_3^\+(\k)}{N!}
 \left[\Phi_3^\+(\0)\right]^N|0\>
\end{align}
with the pair creation operator defined by
\begin{align}
 \Phi_3^\+(\0) \equiv \lambda\phi_3^\+(\0)
 + \sum_\p\frac{v_\p}{u_\p}\psi_1^\+(\p)\psi_2^\+(-\p).
\end{align}
Here, if one state $\psi_3^\+(\k)\big[\Phi_3^\+(\0)\big]^N|0\>$ is
viewed as the atom-like impurity, then another superposed state
$\big[\psi_3^\+(\k)\Phi_3^\+(\0)\big]\big[\Phi_3^\+(\0)\big]^N|0\>$ can
be viewed as the trimer-like impurity in which the atom-like impurity is
dressed by a zero-momentum pair extracted from the background
condensate (see Fig.~\ref{fig:continuity}).  In particular, because of
the existence of the latter component in the atom-like wave function
(\ref{eq:atom}), it can hybridize with the following trimer-like wave
function, i.e., $\<\A(\k)|H|\T(\k)\>\neq0$:
\begin{align}\label{eq:trimer}
 & |\T(\k)\> = \left[\frac1{\sqrt{V}}\sum_{\sigma=1,2}\sum_\p
 \alpha_{\k\sigma}(\p)\psi_\sigma^\+(\p)\phi_\sigma^\+(\k-\p)\right. \notag\\
 &\quad + \frac1{\sqrt{V}}\sum_{\P\neq\0}
 \psi_3^\+(\k-\P)\,\Bigg\{\beta_\k(\P)\phi_3^\+(\P) \notag\\
 &\quad + \left.\left.\!\frac1{\sqrt{V}}\sum_\p
 \gamma_\k(\P,\p)\psi_1^\+(\p)\psi_2^\+(\P-\p)\right\}\right]|\SF\>.
\end{align}
This trimer-like wave function is constructed so that it becomes the
most general three-body wave function in the zero density limit
$|\SF\>\to|0\>$ and thus is capable of exactly reproducing the trimer
formation in the vacuum.  We note that the sum in the second line of
Eq.~(\ref{eq:trimer}) excludes $\P=\0$ because the corresponding
component already exists in Eq.~(\ref{eq:atom}) as is evident from its
decomposition (\ref{eq:decomposition}) and, consequently, the above two
wave functions have no overlap $\<\A(\k)|\T(\k)\>=0$.  With all these
preparations, we finally propose the ansatz
\begin{align}\label{eq:atom-trimer}
 |\AT(\k)\> = |\A(\k)\> + |\T(\k)\>
\end{align}
as a possible crossover wave function that smoothly interpolates
atom-like $|\A(\k)\>$ and trimer-like $|\T(\k)\>$ impurities in a
superfluid state and refer to it as a polaronic atom-trimer state.

At this point, it is worthwhile to contrast our polaronic atom-trimer
state (\ref{eq:atom-trimer}) with the usual polaron in a two-component
Fermi gas~\cite{Schirotzek:2009,Massignan:2014}.  In the latter, a
single impurity atom is dressed by a pair of particle and hole excited
from the background Fermi sea~\cite{Chevy:2006,Combescot:2007} and thus
distinct from the trimer~\cite{Mathy:2011,Parish:2013}, while in the
former, a single impurity atom is dressed by a pair of two particles
extracted from the background condensate and thus indistinguishable from
the trimer.  A similar situation can be found in Bose-Fermi mixtures
where fermionic atoms and molecules are indistinguishable in the
presence of condensed
bosons~\cite{Powell:2005,Marchetti:2008,Rath:2013}.  In principle, the
single impurity atom can be further dressed by two or more condensed
pairs to form a pentamer-like or larger molecule, but their
contributions to the impurity wave function are expected to be
insignificant because their formations are not favored by the Pauli
exclusion principle between the same component of fermions.  We also
note that our crossover wave function (\ref{eq:atom-trimer}) differs in
the spirit from the ansatz in Ref.~\cite{Rapp:2007} which is intended to
describe the sharp transition from the superfluid phase to the trimer
liquid phase.

\section{Polaron phase diagram}
The variational parameters $z_\k$, $\alpha_{\k\sigma}(\p)$,
$\beta_\k(\P)$, $\gamma_\k(\P,\p)$ in the trial wave function
(\ref{eq:atom-trimer}) are determined so as to minimize its energy
expectation value $\eps_\AT(\k)\equiv\<\AT(\k)|H|\AT(\k)\>-E_\SF$ under
the normalization condition $\<\AT(\k)|\AT(\k)\>=1$.  After
straightforward calculations, we find that
$\alpha_\k(\p)\equiv\alpha_{\k1}(\p)=\alpha_{\k2}(\p)$ and
$\beta_\k(\P)$ simultaneously satisfy
\begin{subequations}\label{eq:integral}
\begin{align}\label{eq:alpha}
 & \left[-\frac{m}{4\pi a} + \frac{m^2R_*}{4\pi}
 \left\{\frac{\eps_{\k-\p}}2-\mu+\Ep-\eps_\AT(\k)\right\}\right. \notag\\
 & - \left.\int\!\frac{d\P}{(2\pi)^3}\left\{\frac{|u_{\P-\p}|^2}
 {\eps_{\k-\P}+\Ep+E_{\P-\p}-\eps_\AT(\k)}-\frac1{2\ep}\right\}\right]
 \alpha_\k(\p) \notag\\
 &= \frac{\Ep-\ep+\mu}{\ek-\eps_\AT(\k)}
 \int\!\frac{d\q}{(2\pi)^3}\frac{\alpha_\k(\q)}{\Eq} \notag\\
 &\quad + \int\!\frac{d\P}{(2\pi)^3}\frac{|u_{\P-\p}|^2
 \left[\alpha_\k(\P-\p)+\beta_\k(\P)\right]}{\eps_{\k-\P}+\Ep+E_{\P-\p}-\eps_\AT(\k)}
\end{align} 
and
\begin{align}
 & \left[-\frac{m}{4\pi a} + \frac{m^2R_*}{4\pi}
 \left\{\eps_{\k-\P}+\frac\eP2-2\mu-\eps_\AT(\k)\right\}\right. \notag\\
 & - \left.\int\!\frac{d\p}{(2\pi)^3}\left\{\frac{|u_\p|^2|u_{\P-\p}|^2}
 {\eps_{\k-\P}+\Ep+E_{\P-\p}-\eps_\AT(\k)}-\frac1{2\ep}\right\}\right]
 \beta_\k(\P) \notag\\
 &= 2\int\!\frac{d\p}{(2\pi)^3}\frac{|u_\p|^2|u_{\P-\p}|^2
 \alpha_\k(\p)}{\eps_{\k-\P}+\Ep+E_{\P-\p}-\eps_\AT(\k)}.
\end{align}
\end{subequations}
These two coupled integral equations have similar structures to the
Skorniakov--Ter-Martirosian-type equations to determine the trimer
binding energy in the vacuum~\cite{Nishida:2012}, while the first term
in the right-hand side of Eq.~(\ref{eq:alpha}) is the exception
originating from the hybridization between the atom-like and trimer-like
wave functions $\<\A(\k)|H|\T(\k)\>\neq0$ and thus plays an essential
role here.  By solving Eqs.~(\ref{eq:mean-field}) and
(\ref{eq:integral}) together with the normalization condition for a
given set of inverse scattering length $1/a\kF$ and resonance range
$R_*\kF$~\cite{normalization}, we obtain the quasiparticle energy
$\eps_\AT(\k)$ and the normalized wave function $|\AT(\k)\>$ of the
polaronic atom-trimer state (\ref{eq:atom-trimer}).  In particular, the
important quantity of our interest is the quasiparticle weight
$|z_\k|^2=\<\A(\k)|\AT(\k)\>$, which measures the atomic fraction
contained in the impurity wave function while the rest
$1-|z_\k|^2=\<\T(\k)|\AT(\k)\>$ measures the trimeric fraction therein.
Therefore, it can serve as an indicator of the polaronic atom-trimer
continuity.

\begin{figure}[t]
 \includegraphics[width=\columnwidth]{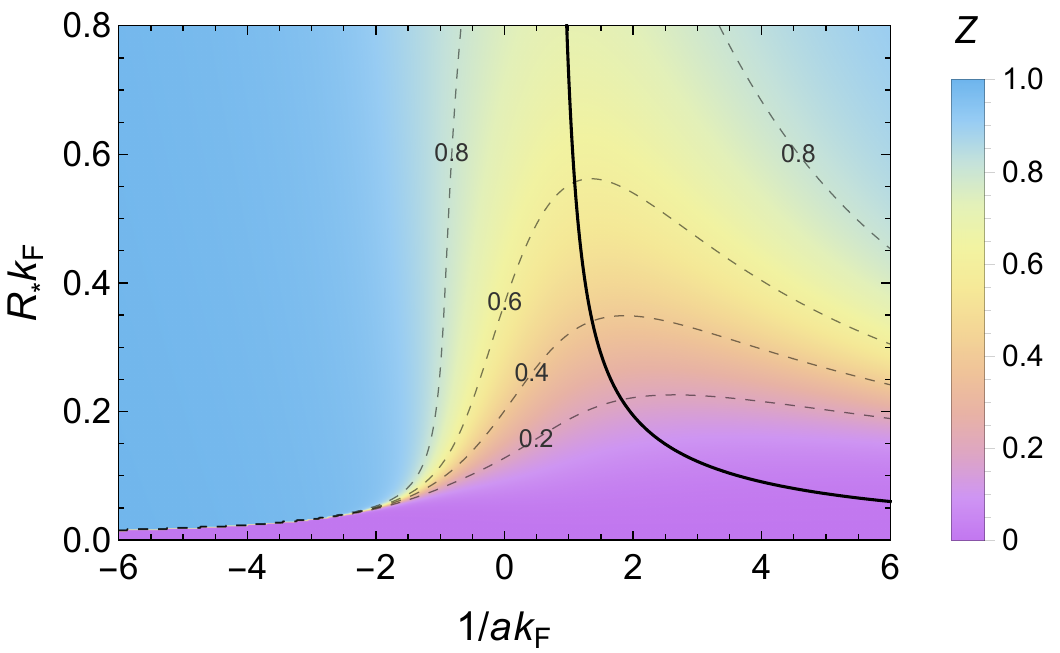}
 \caption{Atomic fraction $Z\equiv|z_{\k=\0}|^2$ in the polaronic
 atom-trimer state (\ref{eq:atom-trimer}) in the plane of the inverse
 scattering length $1/a\kF$ and the resonance range $R_*\kF$.  The
 impurity state is atom-like in the blue region ($Z\gtrsim0.8$) while
 trimer-like in the purple region ($Z\lesssim0.2$) and they are smoothly
 connected in the yellow-to-orange crossover region
 ($0.2\lesssim Z\lesssim0.8$).  The black solid curve indicates the
 phase boundary above which the dimer state (\ref{eq:dimer}) has the
 lower energy than the polaronic atom-trimer state.
 \label{fig:phase-diagram}}
\end{figure}

Figure~\ref{fig:phase-diagram} shows the numerically obtained atomic
fraction $Z\equiv|z_\0|^2$ assuming zero center-of-mass momentum $\k=\0$
as well as zero orbital angular momentum, i.e.,
$\alpha_\0(\p)=\alpha_\0(|\p|)$ and $\beta_\0(\P)=\beta_\0(|\P|)$, where
the polaronic atom-trimer state is expected to have the lowest energy.
In the plane of $1/a\kF$ and $R_*\kF$, we find that the impurity state
is atom-like $Z\sim1$ in the weak coupling (small $1/a\kF$) or narrow
resonance (large $R_*\kF$) region while it becomes trimer-like $Z\sim0$
toward the strong coupling (large $1/a\kF$) and broad resonance (small
$R_*\kF$) region.  These two regions are indeed smoothly connected in
between and thus exhibit the polaronic atom-trimer continuity.  We note
that this smooth crossover from the atom-like impurity to the
trimer-like impurity becomes increasingly sharper toward the weak
coupling and broad resonance limit, $1/a\kF\to-\infty$ and $R_*\kF\to0$,
which is equivalent to taking the zero density limit $\kF\to0$ and thus
the phase boundary there coincides with the threshold $R_*/a=-0.0917249$
for the trimer formation in the vacuum~\cite{Nishida:2012}.

\begin{figure}[t]
 \includegraphics[width=0.9\columnwidth]{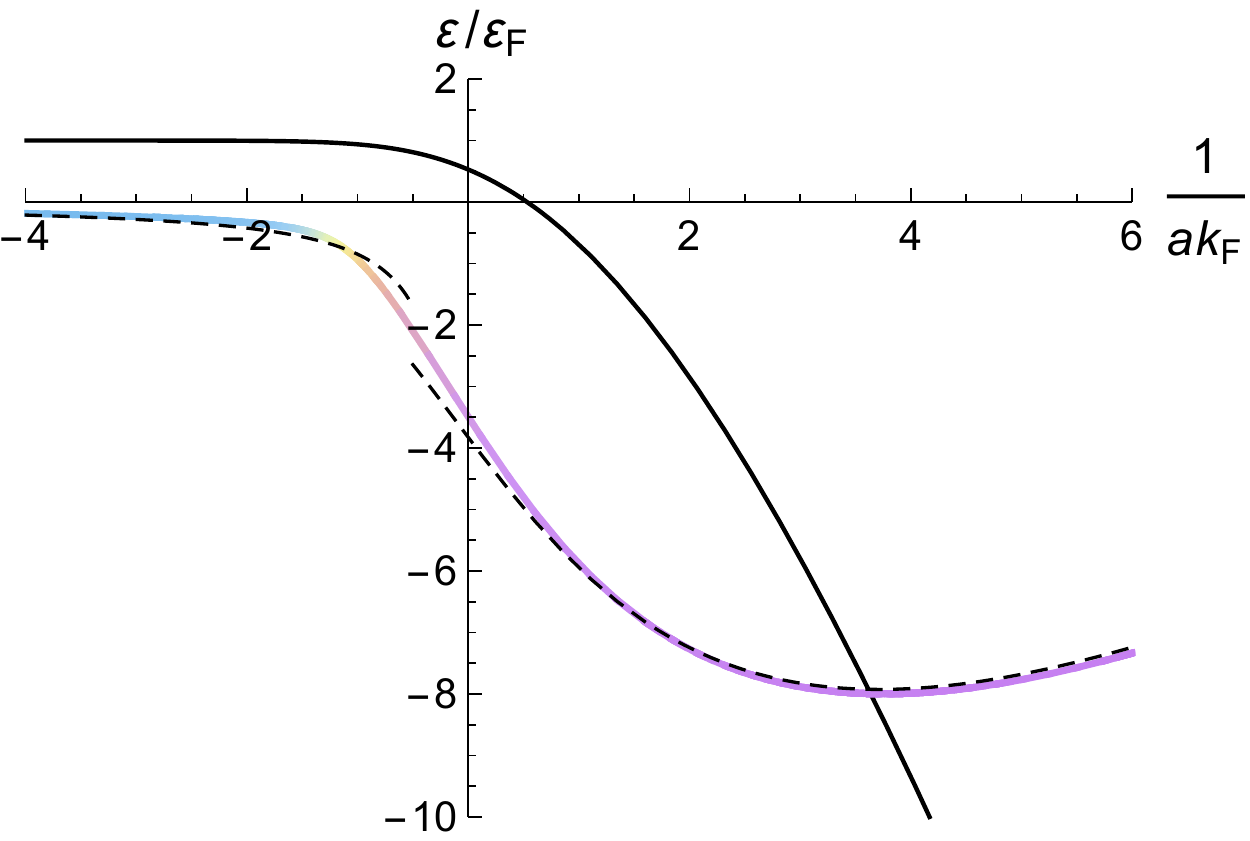}
 \caption{Polaronic atom-trimer energy $\eps_\AT(\k=\0)$ in units of the
 Fermi energy $\eF\equiv\kF^2/(2m)$ at $R_*\kF=0.1$ as a function of
 $1/a\kF$.  It is plotted by the colored solid curve with the atomic
 fraction $Z$ indicated by the same color as in
 Fig.~\ref{fig:phase-diagram}.  The upper left dashed curve is the
 mean-field polaron energy $\eps=(4\pi a/m)(\kF^3/3\pi^2)$
 [Eq.~(\ref{eq:polaron})] while the lower right one is the vacuum trimer
 binding energy $\eps=E_3-2\mu$ [Eq.~(\ref{eq:vacuum})].  The black
 solid curve represents the twofold degenerate dimer energy
 $\eps_\D(\0)=\mu$ [Eq.~(\ref{eq:chemical})] obtained from the simple
 ansatz (\ref{eq:dimer}).  \label{fig:energy}}
\end{figure}

The polaronic atom-trimer continuity is also exhibited by the
quasiparticle energy $\eps_\AT(\0)$ of the polaronic atom-trimer state
(\ref{eq:atom-trimer}).  In the weak coupling limit $1/a\kF\to-\infty$
where the impurity state is atom-like $Z\to1$, its energy resulting from
Eqs.~(\ref{eq:integral}) reduces to the usual mean-field polaron
energy~\cite{Combescot:2007}
\begin{align}\label{eq:polaron}
 \eps_\AT(\0) \to \frac{4\pi a}{m}\left(n_1+n_2\right).
\end{align}
On the other hand, in the broad resonance limit $R_*\kF\to0$ where the
impurity state is trimer-like $Z\to0$, its energy resulting from
Eqs.~(\ref{eq:integral}) can be expressed in terms of the trimer binding
energy $E_3$ in the vacuum~\cite{binding} as
\begin{align}\label{eq:vacuum}
 \eps_\AT(\0) \to E_3-2\mu.
\end{align}
These two asymptotic formulas as well as the numerically obtained
polaronic atom-trimer energy $\eps_\AT(\0)$ are plotted in
Fig.~\ref{fig:energy} as functions of $1/a\kF$ by choosing $R_*\kF=0.1$
for demonstration.  Here we find that the mean-field polaron energy
(\ref{eq:polaron}) valid in the atom-like region ($1/a\kF\lesssim-1$)
and the vacuum trimer binding energy (\ref{eq:vacuum}) valid in the
trimer-like region ($1/a\kF\gtrsim0$) are indeed smoothly interpolated
by the polaronic atom-trimer energy $\eps_\AT(\0)$ and thus we again
confirm the polaronic atom-trimer continuity.

So far we have considered that a single impurity atom of the third
component binds two fermions from a superfluid state of the other two
components to form a trimer-like impurity.  On the other hand, there
exists another possibility in which the single impurity atom binds one
fermion to form a dimer-like impurity and becomes distinct from the
above polaronic atom-trimer state (see Fig.~\ref{fig:continuity}).  Such
a dimer state is twofold degenerate with respect to the exchange of
components $1\leftrightarrow2$ and the simplest trial wave function to
describe one of the two is
\begin{align}\label{eq:dimer}
 |\D(\k)\> = \left[\bar\beta_\k\phi_1^\+(\k)
 + \frac1{\sqrt{V}}\sum_\p\bar\gamma_\k(\p)\psi_2^\+(\p)\psi_3^\+(\k-\p)\right]|\SF\>.
\end{align}
By minimizing its energy expectation value
$\eps_\D(\k)\equiv\<\D(\k)|H|\D(\k)\>-E_\SF$ with respect to the
variational parameters $\bar\beta_\k$, $\bar\gamma_\k(\p)$ under the
normalization condition $\<\D(\k)|\D(\k)\>=1$~\cite{solution}, we obtain
an equation solved by $\eps_\D(\k)$:
\begin{align}\label{eq:energy}
 & -\frac{m}{4\pi a}
 + \frac{m^2R_*}{4\pi}\left[\frac\ek2-\mu-\eps_\D(\k)\right] \notag\\
 &= \int\!\frac{d\p}{(2\pi)^3}
 \left[\frac{|u_\p|^2}{\Ep+\eps_{\k-\p}-\eps_\D(\k)}-\frac1{2\ep}\right].
\end{align}
In particular, by comparing Eq.~(\ref{eq:energy}) with the gap equation
(\ref{eq:gap}), its solution at zero center-of-mass momentum $\k=\0$ is
easily identified as
\begin{align}\label{eq:chemical}
 \eps_\D(\0) = \mu.
\end{align}
The resulting dimer energy at $R_*\kF=0.1$ is also plotted in
Fig.~\ref{fig:energy} where we find that the twofold degenerate dimer
states have the lower energy than the polaronic atom-trimer state in the
strong coupling region $1/a\kF>3.6$.  Therefore, in addition to the
above smooth crossover from the atom-like impurity to the trimer-like
impurity, there is a sharp transition to the dimer-like impurity with
increasing the attraction $1/a\kF$ and the corresponding phase boundary
in the plane of $1/a\kF$ and $R_*\kF$ is indicated in
Fig.~\ref{fig:phase-diagram}.  This sharp transition from the polaronic
atom-trimer state (\ref{eq:atom-trimer}) to the dimer state
(\ref{eq:dimer}) is indeed a close analog of the polaron-molecule
transition in a two-component Fermi gas~\cite{Prokof'ev:2008}.

\section{Summary and outlook}
This Letter is aimed at developing a new crossover theory that smoothly
interpolates atom and trimer in three-component Fermi gases.  To this
end, we took a variational approach in the vicinity of a narrow Feshbach
resonance and successfully showed that a single impurity atom in the
presence of condensed pairs smoothly changes its character from atom to
trimer with increasing the attraction and eventually there is a sharp
transition to dimer.  The emergent polaronic atom-trimer continuity is
signaled by the rapid decrease of the quasiparticle weight $|z_\k|^2$ as
well as the polaron energy $\eps_\AT(\k)$ evolving into the trimer
binding energy (see Fig.~\ref{fig:energy} for $\k=\0$).  These key
quantities are actually measurable in ultracold atom experiments by
transferring an impurity atom with momentum $\k$ from a non-interacting
state to an interacting state, whose transition rate at frequency
$\omega$ exhibits a quasiparticle peak
of~\cite{Massignan:2011,Schmidt:2011,Schmidt:2012,Shashi:2014}
\begin{align}
 I_\k(\omega)|_\mathrm{peak} = |z_\k|^2\delta[\omega+\ek-\eps_\AT(\k)].
\end{align}
Therefore, by utilizing this inverse radio-frequency or microwave
spectroscopy~\cite{Frohlich:2011,Kohstall:2012}, the polaronic
atom-trimer continuity may be experimentally probed, for example, with a
three-component Fermi gas of
$^6$Li~\cite{Ottenstein:2008,Huckans:2009,Nakajima:2010} or a recently
realized superfluid Bose-Fermi mixture of
$^7$Li-$^6$Li~\cite{Ferrier-Barbut:2014}.

Our findings on the polaron problem also have immediate consequences on
the phase diagram of a three-component Fermi gas with a small
concentration $n_3\ll n_1=n_2$ introduced to the third component of
fermions.  When the polaronic atom-trimer state (\ref{eq:atom-trimer})
is the ground state, it will form an unpaired Fermi sea coexisting with
the condensed pairs of the other two components.  Because its many-body
wave function is provided by
\begin{align}
 |\FS\>_\AT = \prod_{\eps_\AT(\k)<\mu_3}\Psi_3^\+(\k)|\SF\>
\end{align}
with $\Psi_3^\+(\k)$ being the creation operator of a single atom-trimer
state $\Psi_3^\+(\k)|\SF\>\equiv|\AT(\k)\>$, the unpaired Fermi sea of
atoms smoothly evolves into that of trimers with
Fig.~\ref{fig:phase-diagram} scanned from the upper left region to the
lower region.  On the other hand, when the twofold degenerate dimer
states (\ref{eq:dimer}) are the ground state, they will undergo the
Bose-Einstein condensation and thus the three different pair condensates
coexist with no unpaired fermions in the upper right side of the phase
boundary in Fig.~\ref{fig:phase-diagram}.  It is impressive to find that
the polaron phase diagram obtained here is already similar in essential
features to the schematic phase diagram proposed in the case of equal
populations for all three components of fermions~\cite{Nishida:2012}.
Our novel crossover wave function (\ref{eq:atom-trimer}) properly
incorporating the polaronic atom-trimer continuity, as well as the dimer
wave function (\ref{eq:dimer}), will provide a useful basis to further
investigate the phase diagram of three-component Fermi gases in more
general situations, including unequal masses, populations, and
inter-component interactions, relevant to ultracold atom
experiments~\cite{Ottenstein:2008,Huckans:2009,Nakajima:2010,Ferrier-Barbut:2014}.

\acknowledgments
This work was supported by JSPS KAKENHI Grant Number 25887020.  Part of
the numerical calculations were carried out at the YITP computer
facility in Kyoto University.


\begin{thebibliography}{99}
\bibitem{normalization}
Once the solutions $\alpha_\k(\p)$ and $\beta_\k(\P)$ are obtained from
Eqs.~(\ref{eq:integral}), the other two parameters are determined by
\begin{align*}
 z_\k = \frac{g\Delta^*}{\ek-\eps_\AT(\k)}
 \int\!\frac{d\q}{(2\pi)^3}\frac{\alpha_\k(\q)}{\Eq}
\end{align*}
and
\begin{align*}
 \gamma_\k(\P,\p) = g\,\frac{\alpha_\k(\p)+\alpha_\k(\P-\p)+\beta_\k(\P)}
 {\eps_{\k-\P}+\Ep+E_{\P-\p}-\eps_\AT(\k)},
\end{align*}
which have to be normalized according to
\begin{align*}
 |z_\k|^2 & + 2\int\!\frac{d\p}{(2\pi)^3}|u_\p|^2|\alpha_\k(\p)|^2
 + \int\!\frac{d\P}{(2\pi)^3}|\beta_\k(\P)|^2 \notag\\
 & + \iint\frac{d\P d\p}{(2\pi)^6}\,|u_\p|^2|u_{\P-\p}|^2|\gamma_\k(\P,\p)|^2 = 1.
\end{align*}
\bibitem{solution}
The normalized solutions are found to be
\begin{align*}
 |\bar\beta_\k|^2 = \left[1 + g^2\int\!\frac{d\p}{(2\pi)^3}
 \frac{|u_\p|^2}{[\Ep+\eps_{\k-\p}-\eps_\D(\k)]^2}\right]^{-1}
\end{align*}
and $\bar\gamma_\k(\p)=g\bar\beta_\k/[\Ep+\eps_{\k-\p}-\eps_\D(\k)]$.
\end{thebibliography}
\end{document}